\documentstyle[pre,aps]{revtex}

\newcommand{\be}{\begin{equation}}
\newcommand{\ee}{\end{equation}}
\newcommand{\ra}{\rightarrow}
\newcommand{\al}{\alpha}
\newcommand{\pr}{\partial}
\newcommand{\sr}{\stackrel{\rightarrow}{r}}
\newcommand{\dps}{\psi^\dagger_{\nu s}}
\newcommand{\dc}{c^\dagger}
\newcommand{\ds}{\sigma^\dagger}
\newcommand{\dpss}{\psi^\dagger_{\nu s'}}
\newcommand{\sk}{\stackrel{\rightarrow}{k}}
\newcommand{\spp}{\stackrel{\rightarrow}{p}}
\newcommand{\g}{\stackrel{-}{\Gamma}_1(\sk_1,\sk_2,\sk_3,\sk_4)}
\newcommand{\gn}{\stackrel{-}{\Gamma}_\nu (\sk_1,\sk_2,\sk_3,\sk_4)}
\newcommand{\ep}{\varepsilon}
\newcommand{\srr}{\stackrel{\rightarrow}{r}}

\begin{document}

\draft

\title{SUPERCONDUCTORS WITH MESOSCOPIC PHASE SEPARATION}
\author{A. J. Coleman, E. P. Yukalova and V. I. Yukalov} 
\address{\it Department of Mathematics and Statistics \\ 
Queen's University, Kingston, Ontario K7L 3N6, Canada}

\maketitle

\begin{abstract}

A model of superconductivity is proposed taking into account repulsive 
particle interaction, mesoscopic phase separation and softening of crystalline
lattice. These features are typical of many high--temperature superconductors.
The main results obtained for the model are: (i) phase separation is possible 
only if repulsive forces play a significant role; (ii) the critical 
temperature as a function of the superconducting phase fraction can have 
non--monotonic behaviour; (iii) superconductivity is possible in heterophase 
systems even when it would be forbidden in pure samples. These results are in 
agreement with experiments.
\end{abstract}

\section{Introduction}

In the standard theory of superconductivity one employs a reduced Hamiltonian 
involving only the attractive part of an effective particle interaction, 
responsible for the existence of superconductivity, and omitting the repulsive
part of the interaction as irrelevant. This approach was adapted in the 
original paper of Bardeen, Cooper and Schrieffer [1]. The consideration of 
the Coulomb interaction leads to a renormalization of the coupling constant 
[1,2], which can be effectively included in the formula for the 
superconducting critical temperature [3]. In usual metals the averaged 
Coulomb interaction is small as compared to the coupling parameter mediated 
by phonon exchange. A detailed discussion of reasons why the repulsive 
interaction may be neglected has been given in a review by Legget [4].

These arguments concern the {\it direct} influence of repulsive interactions. 
However, they can affect superconductivity {\it indirectly}. An obvious 
example is that all interactions are included in the definition of the 
electronic energy bands and the corresponding density of states [5,6]. An 
interaction that is not directly responsible for the pairing mechanism, can, 
nevertheless, drastically strengthen the order by providing a high density of 
states at the Fermi surface, thus ensuring a large value of the effective 
coupling parameter, which, in turn, leads to a higher superconducting 
transition temperature. Such an increase of the density of states can occur, 
for instance, when a van Hove singularity is located almost precisely at the 
Fermi level [7].

It is necessary to keep in mind that it is Coulomb interactions, and not 
pairing interactions that define the space structure of matter. The electrical
properties change drastically from one structure to another [5-8]. Moreover, 
the stability or instability of a crystalline (and chemical) structure is 
closely connected with superconductivity. This was noticed long before the 
discovery of the high - temperature superconductors. For example, Testardi 
[9] analysed a large number of superconducting A-15 compounds and concluded 
that superconductivity and structural instability are really related: the 
higher the superconducting critical temperature, the more probable is the 
occurrence of structural instability. Indeed, most superconductors with 
$ \; T_c \sim 20K \; $ clearly exhibit instability. 

The observation that higher superconducting $ \; T_c \; $ is almost always 
related to increased instability received strong confirmation after the 
historic discovery of high -$ \; T_c \; $ superconductors by Bednorz and 
M\"uller [10]. Possibly, the most careful, critical, material systematics for 
high -$ \; T_c \; $ superconductors has been given by Phillips [11]. He 
stresses that if we review the known high -$ \; T_c \; $ superconducting 
materials we are unavoidably confronted with mechanical instabilities as a 
factor which always accompanies $ \; T_c \; $, whether in intermetallic or 
oxide - based superconductors. This occurs so often that we would hard put 
to dismiss the repeated appearance of such instabilities as mere coincidence. 
According to Phillips [11], there is no room for doubt that it is lattice 
instabilities (and nothing else) that produce high -$ \; T_c \; $ 
superconductivity just as much in cuprates as in intermetallic compounds.

The main result of structural instabilities is an essential softening of the 
lattice. This softening is manifested in various precursor effects occurring 
near the superconducting transition temperature, such as anomalies in elastic 
moduli, in strain parameters, and in sound velocity [9,11]. The softening of 
phonon modes can be observed by infrared, Raman, and neutron scattering 
experiments [11]. Nuclear gamma - resonance studies [12,13] also show that a 
dramatic lattice softening occurs at $ \; T_c \; $ for high - temperature 
superconductors, the most noticable result of which is an anomalous sagging 
of the M\"ossbauer factor.

The review of experimental data also provides evidence that the lattice 
softening and the increase in $ \; T_c \; $ are both associated with the 
formation of phase precursor clusters with structural (spatial or chemical) 
composition different from the basic component [9,11]; the linear size of the 
clusters is of the order $ \; 10 - 100 \AA \; $. Such mixed structures exist 
in many ceramic superconductors, as has been observed by using pulsed--neutron
scattering, synchrotron $ \; x \; $- ray powder diffraction, nuclear magnetic 
resonance, nuclear quadrupole resonance, and nuclear gamma resonance 
[5,11,14]. For instance, the low - temperature orthorhombic and 
high--temperature tetragonal structures coexist in oxide superconductors in a 
region around  $ \; T_c \; $ [15-17]. A more detailed discussion of these and 
other experiments can be found in a recent review [18]. The relation between 
the existence of soft vibrational modes and the appearance of clusters of a 
mesoscopic size has been also proved by molecular dynamics computer simulation
for metastable glassy models [19,20].

The clusters of one phase structure inside another, occurring near a phase 
transition point, are randomly distributed in space, and often, being 
unstable, change with time. To emphasize the randomness of their distribution,
they can be called {\it heterostructural}, or {\it heterophase fluctuations}, 
and to stress their mesoscopic sizes, they can be named {\it mesoscopic 
structural fluctuations} [18].

Since structural characteristics are intimately related to electronic and 
conducting properties, the existence of structural fluctuations implies the 
appearance of spatial fluctuations in the superconducting properties. 
Mesoscopic structural fluctuations are accompanied by large energy 
fluctuations which favour a stochastic separation of a sample into 
superconducting and normal regions [18,21]. There is clear experimental 
evidence that the superconducting cuprates display a nanoscale phase 
separation into insulating and superconducting nanodomains [7]. This is in 
agreement with some simple models [7,22-24] predicting that such a phase 
separation can be thermodynamically profitable. Recent experiments with 
high - temperature superconductors confirm that only part of a given 
sample - often less than a few percent - is in a superconducting phase 
(see [25-27] and references therein).

In the present paper we suggest a way of taking into account the three 
mutually interrelated factors: Coulomb interaction, phase separation, and 
lattice softening; all of which factors are very important in 
high - temperature superconductors. To clearly emphasize the influence of 
these factors as such we shall use the standard approximations accepted for 
superconductors. We give a detailed analysis of the superconducting critical 
temperature as a function of parameters related to the attracting and 
repulsive interactions and to the superconducting phase fraction.

Everywhere below we use the system of units with 
$ \; \hbar \equiv k_B \equiv 1; $.

\section{Phase separation} 

In order to take into account phase separation, we must admit such a 
possibility from the beginning, and then the corresponding phase probabilities
must be defined in a self - consistent way. Assume that a phase separation has
occurred so that our sample consists of regions occupied by two different 
phases numbered by the index $ \; \nu = 1,2 \; $; the phase regions being 
randomly distributed in space. Assign the index $ \; \nu = 1 \; $ to the 
superconducting phase, and $ \; \nu = 2 \; $, to the normal phase. For each 
phase distribution, or phase configuration, the sample is nonuniform, which 
greatly complicates the consideration. However, assuming that each phase 
distribution is random, we may average observable quantities over these phase 
configurations. As a result, we come to a system described by a renormalized 
Hamiltonian that allows usual techniques supplemented with additional 
equations for phase probabilities and conditions of stability. The procedure 
of averaging over configurations has been expounded, in all necessary detail, 
in earlier papers [28-31] and discussed in a recent review [18]. Therefore, 
we shall not repeat it here, especially since the mathematical foundation is 
quite long and tedious even though the final result is rather simple.

The averaging over phase configurations, in the case of a random mixture of 
two phases, leads to the definition of a renormalized Hamiltonian
\be
\stackrel{-}{H} = H_1 \oplus H_2 ,
\ee
which is a direct sum of terms
\be
H_\nu = w_\nu H_\nu^{kin} + w_\nu^2H_\nu^{int} ,
\ee
in which $ \; H_\nu^{kin} \; $ is a kinetic - energy operator, including 
external fields, if any, $ \; H_\nu^{int} \; $ is an interaction - energy 
operator. The renormalizing factor $ \; w_\nu \; $ is the geometric 
probability of the $ \; \nu \; $- phase, that is, the ratio of the average 
volume occupied by the $ \; \nu \; $- phase to the total volume of the sample.
The Hamiltonian (2) is defined on the space
\be
{\cal Y} = {\cal Y}_1 \otimes {\cal Y}_2 ,
\ee
which is a tensor product of Hilbert spaces, where $ \; {\cal Y}_\nu \; $ is 
the space of states typical of the $ \; \nu \; $- phase. When $\;\nu = 1\;$ 
corresponds to a superconducting phase, and $ \; \nu = 2 \; $, to a normal 
phase, the state - spaces are chosen so that the so - called anomalous 
averages calculated with the states from $ \; {\cal Y}_1 \; $ are not 
identically zero, while those calculated with the states from $\;{\cal Y}_2\;$
are zero. In other words, the gaps related to the anom
alous averages satisfy the condition
\be
\Delta_1 \not\equiv 0 , \qquad \Delta_2 \equiv 0 .
\ee
This condition can be interpreted in several ways. For example, we can always 
posit that each Hilbert space of microstates, $ \; {\cal Y}_\nu \; $, is 
restricted to functions which are invariant under a symmetry group yielding 
desired properties for the order parameters [32]. Then, the order parameters 
are just the gaps in (4), and the corresponding microstates can, for example, 
be chosen as the BCS wave functions [1,33,34] for the superconducting phase 
and the usual Slater determinants for the normal phase. Another way to 
interpret the choice of the order parameters (4) is to connect them, and the 
values of the anomalous averages, with the off - diagonal long - range order 
of reduced density matrices [35]. The largest eigenvalues of the latter are 
also known [35-37] to be directly related to this off - diagonal order. A more
detailed discussion of these relations can be found in reviews  [38,39]. Such 
relations between microscopic and macroscopic characteristics can be described
in a general way by introducing the notion of order indices [40,41]. If 
$ \; {\hat \rho}_n^N \; $ is an $ \; n \; $- particle reduced density matrix 
for a system of $ \; N \; $ particles, then the order indices are defined as
$$ \alpha_n \equiv 
\lim_{N \ra \infty}\frac{\log || {\hat \rho}_n^N ||}{\log N} . $$
It has been shown [41] that the index $ \; \alpha_n \; $ satisfies the 
restriction
$$ 0 \leq \alpha_n \leq n \qquad ( 0 \leq n \leq N ) . $$
Different thermodynamic phases correspond to different sets of order indices, 
so that a $ \; \nu \; $- phase is characterized by a set 
$ \{ \alpha_n^{(\nu )} \} \; $. Necessary and sufficient conditions that the 
phase with $ \; \nu =1 \; $ is superconducting and the phase with $\;\nu=2\;$
is normal are given by
\be
\al_{2n}^{(1)} = n , \qquad \al_{2n}^{(2)} < n .
\ee
Thus, there are several well developed methods for defining thermodynamic 
phases. What we still need to define is how to find the geometric 
probabilities $ \; w_\nu \; $ of the corresponding phases when considering a 
mixture described by the Hamiltonian (1).

According to the general procedure [18], the phase probabilities are given by 
the minimization of the thermodynamic potential
\be
f = -\frac{T}{N}\ln Tr e^{-\beta \stackrel{-}{H}} \qquad (\beta T \equiv 1 )
\ee
under the condition
\be
w_1 + w_2 = 1 , \qquad 0 \leq w_\nu \leq 1 ,
\ee
where $ \; T \; $ is temperature. The normalization condition (7) can be taken
into account explicitly by using the notation
\be
w_1 \equiv w \qquad w_2 \equiv 1 - w .
\ee
Then $ \; w \; $ satisfies the equations
\be
\frac{\pr f}{\pr w} = 0 , \qquad \frac{\pr^2 f}{\pr w^2} > 0 .
\ee 
The first of the equations in (9) is equivalent to
\be
\langle \frac{\pr \stackrel{-}{H}}{\pr w} \rangle = 0 .
\ee
Substituting the hamiltonian (1) and introducing the notation
\be
K_\nu \equiv \langle H_\nu^{kin} \rangle , \qquad U_\nu \equiv 
\langle H_\nu^{int} \rangle ,
\ee
we obtain the equation
\be
w = \frac{2U_2+K_2 - K_1}{2(U_1 + U_2)}
\ee
for the probability of the superconducting phase. In case of thermodynamic 
phases with equal average densities, the probabilities $ \; w_\nu \; $ 
coincide with the phase concentrations defined as the ratios of the number of 
particles in each phase to the total number of particles. The average density 
of the superconducting and of the normal phase are practically the same. 
Therefore we may speak of (12) as of the superconducting phase concentration.

The inequality in (9) shows when the phase separation in a sample is 
thermodynamically advantageous, as compared to a pure superconducting system. 
Taking account of (6), we find
\be
\left ( \langle \frac{\pr^2 \stackrel{-}{H}}{\pr w^2} \rangle \right ) > \beta
\langle \left ( \frac{\pr \stackrel{-}{H}}{\pr w} \right )^2 \rangle .
\ee

In accordance with (7) and (12), we have
\be
-2U_1 \leq K_1 - K_2 \leq 2U_2 .
\ee
Substituting the Hamiltonian $ \; \stackrel{-}{H} \; $, defined in (1) and 
(2), into the left - hand side of (13), we get
$$ \langle \frac{\pr^2 \stackrel{-}{H}}{\pr w^2} \rangle = 2(U_1 + U_2 ) . $$
Using this and noticing that the right - hand side of (13) is always 
non - negative, we obtain the inequality
\be
U_1 + U_2 > 0 .
\ee
Thus, phase separation is possible only when there are sufficiently strong 
repulsive interactions in the system for (15) to be satisfied. In other words,
inequality (15) is a {\it necessary condition for phase separation}. This 
general result is in agreement with Hubbard - model calculations for the 
copper - oxide superconductors [42]. According to these it is necessary to 
include a sufficiently strong nearest - neighbour Coulomb repulsion in order 
to produce phase separation.

A physical system will remain in a mixed state, with phase separation, as long
as this is thermodynamically favourable compared with pure phases. The 
boundary, in the space of thermodynamic parameters, between mixed and pure 
phases is the set of points at which a necessary condition for phase stability
is violated, - one of the conditions (13) or (14), for example. Experiment 
supports the usual assumption that transformation of a pure phase into a mixed
phase begins with the appearance of nuclei of a competing phase within the 
pure phase. This process can be called {\it nucleation} and the point in the 
phase diagram at which this occurs, a {\it nucleation point}. At such a point,
some thermodynamic and dynamic characteristics may display a singularity. 
This may occur, for example, in the density - density response function which 
is determined by the Fourier transform of the second - order density matrix. 
The behaviour of the thermodynamic and dynamic characteristics in the vicinity
of such a phase transition is a separate problem which does not fall within 
the purview of the present paper. Rather, it is our object to delineate 
features which we regard as basic to any reasonable model of superconductors 
which exhibit mesoscopic phase separation. 

\section{Specification of Hamiltonian}

Let us now specify the Hamiltonian, defined in (1) and (2), keeping in mind 
that $ \; H_1 \; $ corresponds to the superconducting phase; and $\; H_2\;$, 
to the normal phase. This implies that the corresponding order parameters and 
order indices must satisfy conditions (4) and (5), respectively. These 
conditions can also be formulated for anomalous averages defined through the 
field operators, $ \; \psi_{\nu s }(\stackrel{\ra}{r}) \; $, where $ \; s\;$ 
denotes spin and $ \; \stackrel{\ra}{r} \in {\bf R}^3 \; $.

The anomalous averages
\be
\langle \psi_{1s}(\sr )\psi_{1s}(\sr ')\rangle \not\equiv 0
\ee
for the first representation are not zero, while those for the second 
representation vanish,
\be
\langle \psi_{2s}(\sr )\psi_{2s}(\sr ')\rangle \equiv 0 .
\ee
As is obvious, the conditions (16) and (17) are directly related to the 
corresponding properties of symmetry of microscopic states [4,18] or to the 
values (5) of the order indices [40,41].

The kinetic part of Hamiltonian (2) can be written in the form
\be
H_\nu^{kin} =\int \sum_{s} \dps (\sr ) \left [ K_\nu (\sr , \sr ') - \mu 
\delta (\sr - \sr ')\right ]\psi_{\nu s}(\sr ')d\sr d\sr ' ,
\ee
in which $ \; K_\nu (\sr , \sr ') \; $ is a kinetic kernel including external 
scalar fields, if any, and $ \; \mu \; $ is the chemical potential. Take for 
the interaction part of (2) a general expression 
$$ H_\nu^{int} = \frac{1}{2} \int \sum_{ss'} \dps (\sr_1)\dpss (\sr_2)V_\nu 
(\sr_1,\sr_2,\sr_3,\sr_4)\psi_{\nu s'}(\sr_3) \times $$
\be
\times \psi_{\nu s}(\sr_4) d\sr_1d\sr_2d\sr_3d\sr_4 , 
\ee
where $ \; V(\ldots ) \; $ is an interacting vertex including all effective 
interactions, repulsive and attractive, direct and indirect. For this moment, 
we do not need to specify these interactions. However, as an example, we can 
think of direct interaction as a repulsive Coulomb force taking screening into
account, and that indirect interactions are mediated by an exchange of phonons
or other bosons.

In order not to complicate our discussion with the consequences of a variety 
of other known factors on the properties of superconductors, we limit 
ourselves to isotropic matter. Thus, we leave aside anisotropy effects and 
the related van Hove singularities [7]. For an isotropic system we can use 
the expansion
\be
\psi_{\nu s}(\sr ) =\frac{1}{\sqrt{V}} \sum_{k}c_{\nu s}(\sr ) e^{i\sk\sr} ,
\ee
in which $ \; V \; $ is the total volume of the system.

In what follows, to simplify the notation, we shall accept the convention
\be
c_{1s}(\sk ) \equiv c_s(\sk ) .
\ee
Then, we shall deal mainly with expressions corresponding to the 
superconducting phase, since an identical treatment can be given for the 
normal phase. The difference between the phases can be taken into account at 
the final stage by invoking conditions (16) and (17).

Substituting (20) in (18), and using the property
\begin{eqnarray}
\frac{1}{V}\int e^{i(\sk - \sk ')\sr}d\sr = \delta_{kk'} \equiv \left \{
\begin{array}{cc}
1, & \sk =\sk ' \\ \nonumber
0, & \sk \neq \sk ' , 
\end{array} \right.
\end{eqnarray}
we have
\be
H_1^{kin} =\sum_{k,k',s} \left [ T_1(\sk ,\sk ') - \mu\delta_{kk'}
\right ]
\dc_s(\sk ) c_s(\sk ') ,
\ee
where the convention (21) is used and
\be
T_\nu (\sk ,\sk ') \equiv 
\frac{1}{V}\int e^{-i\sk\sr}K_\nu (\sr , \sr ')e^{i\sk '\sr '}d\sr d\sr ' 
\ee
is the transport matrix. For (19) we get
\be
H_1^{int} =\frac{1}{2V}\sum_{k_1\ldots k_4}\sum_{ss'}
\Gamma_1(\sk_1,\sk_2,\sk_3,\sk_4)
\dc_s(\sk_1)\dc_{s'}(\sk_2)c_{s'}(\sk_3)c_s(\sk_4)
\ee
with the vertex
$$ \Gamma_1(\sk_1,\sk_2,\sk_3,\sk_4) \equiv \frac{1}{V} \int
V_1(\sr_1,\sr_2,\sr_3,\sr_4)\times $$
\be
\times \exp \left \{ -i\sk_1\sr_1 -i\sk_2\sr_2 + i\sk_3\sr_3 + i\sk_4\sr_4 
\right \} d\sr_1d\sr_2d\sr_3d\sr_4.
\ee
In the following step to simplify the Hamiltonian, we use a convenient 
approximation which embodies a possible Cooper pairing. This could be done in 
several equivalent ways. We may construct an approximating Hamiltonian [43] 
acquiring the so - called Hartree - Fock - Bogolubov approximation. The latter
is widely known and is successfully used in various applications, including 
such exotic ones as heated rotating nuclei [44-46]. We may follow a 
variational approach as in the Blatt quasi - chemical equilibrium theory [47].
Or we may use the fundamental ansatz for the many - body wave function 
yielding what is called [38,39] the antisymmetrized geminal power. All these 
approaches are equivalent to one another [4,34,48,49]. To our mind, the main 
common idea lying behind these approaches can, in second - quantization 
language, be expressed by the operator approximation
$$ \dc_1\dc_2c_3c_4 \approx \dc_1c_4\langle\dc_2c_3\rangle + 
\langle\dc_1c_4\rangle\dc_2c_3 - 
\langle\dc_1c_4\rangle\langle\dc_2c_3\rangle - $$
$$ -\dc_1c_3\langle\dc_2c_4\rangle - \langle\dc_1c_3\rangle\dc_2c_4 + 
\langle\dc_1c_3\rangle\langle\dc_2c_4\rangle + $$
\be
+ \dc_1\dc_2\langle c_3c_4\rangle + \langle\dc_1\dc_2\rangle c_3c_4 -
\langle\dc_1\dc_2\rangle\langle c_3c_4\rangle ,
\ee
where, for compactness, we write $ \; c_n \; $ instead of $ c_{s_n}(\sk )\;$. 
The right - hand side of (26) may be called the operator antisymmetrized 
geminal power. Averaging (26), we obtain the familiar approximation for the 
correlation function
\be
\langle\dc_1\dc_2c_3c_4\rangle \cong \langle\dc_1c_4\rangle 
\langle\dc_2c_3\rangle - \langle\dc_1c_3\rangle\langle\dc_2c_4\rangle + 
\langle\dc_1\dc_2\rangle\langle c_3c_4\rangle 
\ee
of the Hartree - Fock - Bogolubov form.

Applying (26) to (24), we present the latter as a sum
\be
H_1^{int} = H_1^{nor} + H_1^{sup} + B_1 ,
\ee
in which the first term
$$ H_1^{nor} = \frac{1}{V} \sum_{k_1\ldots k_4} \sum_{ss'} \g \times $$
\be
\times \left [ \dc_s(\sk_1)c_s(\sk_4)\langle\dc_{s'}(\sk_2)c_{s'}(\sk_3)
\rangle - \dc_s(\sk_1)c_{s'}(\sk_3)\langle\dc_{s'}(\sk_2)c_s(\sk_4)
\rangle \right ]
\ee
has the normal Hartree - Fock form with
\be
\gn \equiv \frac{1}{2} \left [ \Gamma_\nu (\sk_1,\sk_2,\sk_3,\sk_4) + 
\Gamma_\nu (\sk_2,\sk_1,\sk_4,\sk_3) \right ] .
\ee
The second term in (28), that is,
$$ H_1^{sup} = \frac{1}{2V} \sum_{k_1\ldots k_4} \sum_{ss'}
\Gamma_1(\sk_1,\sk_2,\sk_3,\sk_4) \times $$
\be
\times \left [ \dc_s(\sk_1)\dc_{s'}(\sk_2) 
\langle c_{s'}(\sk_3)c_s(\sk_4)\rangle + c_s(\sk_1)c_{s'}(\sk_2) 
\langle\dc_{s'}(\sk_3)\dc_s(\sk_4)\rangle \right ] ,
\ee
contains anomalous averages and is, thus, responsible for superconductivity. 
The last term in (28) is
$$ B_1 =\frac{1}{2} \sum_{k_1\ldots k_4} \sum_{ss'} 
\Gamma_1(\sk_1,\sk_2,\sk_3,\sk_4) \times $$
$$ \times \left [ \langle\dc_s(\sk_1)c_{s'}(\sk_3)\rangle 
\langle\dc_{s'}(\sk_2)c_s(\sk_4)\rangle -  
\langle\dc_s(\sk_1)c_s(\sk_4)\rangle \langle\dc_{s'}(\sk_2)c_{s'}(\sk_3)
\rangle \right. - $$
\be
\left. -\langle\dc_s(\sk_1)\dc_{s'}(\sk_2)\rangle 
\langle c_{s'}(\sk_3)c_s(\sk_4)\rangle \right ] .
\ee

The Hamiltonian defined by (1), together with (2), (18) and (19), describes a 
system in which the total momentum and spin are conserved. Therefore
$$ \langle\dc_s(\sk )c_{s'}(\sk ')\rangle = \delta_{kk'}\delta_{ss'}
\langle\dc_s(\sk )c_s(\sk )\rangle , $$
\be
\langle\dc_s(\sk )\dc_{s'}(\sk ')\rangle = \delta_{-kk'}\delta_{-ss'} 
\langle\dc_s(\sk )\dc_{-s}(-\sk )\rangle .
\ee
If we denote the up and down spins as $ \; s = \uparrow ,\downarrow \; $, 
then in (33) for $ \; s = \uparrow \; $ the notation $ \; -s \; $ means 
$ \; \downarrow \; $, and for $ \; s=\downarrow \; , -s \; $ means 
$ \; \uparrow \; $. For the normal average introduce the notation
\be
n_1(\sk ) \equiv \sum_{s}n_{1s}(\sk ) ; \qquad n_{1s}(\sk ) = 
\langle \dc_s(\sk ) c_s(\sk ) \rangle ,
\ee and for the anomalous average,
\be
\sigma_1(\sk ) \equiv \langle c_{-s}(-\sk ) c_s(\sk ) \rangle .
\ee

The function $ \; n_\nu (\sk ) \; $ is a momentum distribution of particles. 
Since the transport matrix (23) and the vertex (25) do not depend on spin, we 
have
\be
n_\nu (\sk ) = 2n_{\nu s}(\sk ) .
\ee
The normal term (29), with the use of (33) and (34), can be written as
\be
H_1^{nor} =\sum_{kk',s} M_1(\sk ,\sk ') \dc_s(\sk ) c_s(\sk ') ,
\ee
where $ \; M_\nu \; $ is a mass operator with kernel
\be
M_\nu (\sk ,\sk ') =\frac{1}{V} \sum_{p} \left [ 
\stackrel{-}{\Gamma}_\nu (\sk ,\stackrel{\ra}{p},\stackrel{\ra}{p},\sk ') -
\frac{1}{2} 
\stackrel{-}{\Gamma}_\nu (\sk ,\stackrel{\ra}{p},\sk ',\stackrel{\ra}{p})
\right ] n_\nu (\stackrel{\ra}{p}) ,
\ee
By using (33) and (35), the superconducting term (31) becomes
$$ H_1^{sup} = \frac{1}{2} \sum_{kk'p}
\stackrel{-}{\Gamma}_1 (\sk ,\sk ',-\stackrel{\ra}{p},\stackrel{\ra}{p})
\times $$
\be
\times \left [ \dc_s(\sk ) \dc_{-s}(\sk ')\sigma_1(\stackrel{\ra}{p}) +
c_s(\sk ) c_{-s}(\sk ')\stackrel{\dagger}{\sigma}_1(\stackrel{\ra}{p}) 
\right ] .
\ee
And the scalar term (32) simplifies to
$$ B_1 = -\frac{1}{2} \sum_{k} M_1(\sk ,\sk ) n_1(\sk ) - $$
\be
-\frac{1}{V} \sum_{kp} 
\Gamma_1 (\sk ,-\sk ,-\stackrel{\ra}{p},\stackrel{\ra}{p})
\stackrel{\dagger}{\sigma}_1(\sk ) \sigma_1(\stackrel{\ra}{p}) .
\ee

Notice that the necessity of an accurate and detailed analysis of all 
transformations presented here is dictated by our main goal which is to 
consider the role of all terms of the Hamiltonian not omitting any of them. 
As we have already shown and shall see in what follows, all these terms, 
repulsive as well as attractive, play a crucial role for superconductors 
with phase separation.

\section {Gap Equation}

The sole approximation we have invoked so far now is the operator geminal 
power form (26), which is equivalent to the Hartree - Fock  - Bogolubov 
approximation or to the antisymmetrized geminal power approach. The structure 
of the resulting Hamiltonian is still too complex to permit specific 
conclusions. Actually, a group theoretical analysis by Ozaki [50] of the 
Hartree - Fock - Bogolubov approximation, enumerated 73 possible ordered 
states, including 26 superconducting states! The latter include 
non - Cooper pairing, when an electron pair has nonzero momentum, and various 
superconducting states coexisting with other nonsuperconducting orders.

To make the problem tractable, we can resort to an approximation that 
restricts the space of microscopic states to a subspace satisfying some 
additional constraints [48,51,52]. In our case a natural such constraint is 
to consider only those quantum states that conserve momenta, that is, when 
the momentum conservation occurs not only on the average, as in (33), but 
for all operator combinations:
$$ \dc_s(\sk )c_s(\sk ') = \delta_{kk'}\dc_s(\sk ) c_s(\sk ) , $$
\be
\dc_s(\sk ) \dc_{-s}(\sk ') = \delta_{-kk'}\dc_s(\sk )\dc_{-s}(-\sk ) .
\ee
Such a restricted space consists of BCS wave functions that are a particular 
kind of antisymmetric geminal power functions [34,53].

The restriction (41) makes it possible to greatly simplify all formulas. To 
this end, let us introduce the single - particle spectrum
\be
\ep_\nu (\sk ) \equiv T_\nu (\sk ,\sk ) ,
\ee
the diagonal mass - operator
\be
M_\nu (\sk ) \equiv M_\nu (\sk ,\sk )
\ee
and an effective interaction
\be
J_\nu (\sk,\spp ) \equiv \Gamma_\nu (\sk ,-\sk ,-\spp ,\spp ) .
\ee  
Taking account of (41) and (42) - (44) we obtain for the kinetic term (22)
\be
H_1^{kin} =\sum_{k,s}\left [ \ep_1(\sk ) -\mu \right ]\dc_s(\sk )c_s(\sk ) ,
\ee
for the normal term (37)
\be
H_1^{nor} =\sum_{k,s}M_1(\sk )\dc_s(\sk )c_s(\sk ) ,
\ee
and for the superconducting term (39)
$$ H_1^{sup} =\frac{1}{2V} \sum_{kp,s} J_1(\sk ,\spp ) 
\left [ \dc_s(\sk )\dc_{-s}(-\sk )\sigma_1(\spp ) +  \right. $$
\be
\left. + c_s(\sk )c_{-s}(-\sk )\stackrel{\dagger}{\sigma}_1(\spp ) \right ] . 
\ee
The scalar term (40) becomes
\be
B_\nu = -\frac{1}{2}\sum_{k} M_\nu (\sk )n_\nu (\sk ) - 
\frac{1}{V}\sum_{kp}J_\nu (\sk ,\spp )\ds_\nu (\sk )\sigma_\nu (\spp ) .
\ee
Finally, collecting all these terms, for the Hamiltonian of the 
superconducting phase, given in (2), we obtain the expression
\be
H_1 =w_1\sum_{k}\sum_{s}\omega_1 (\sk )\dc_s(\sk )c_s(\sk ) +
w_1^2H_1^{sup} + w_1^2B_1 ,
\ee
in which $ \; H_1^{sup} \; $ and $ \; B_1 \; $ are defined by (47) and 
(48), and
\be
\omega_\nu (\sk ) \equiv \ep_\nu (\sk ) - \mu + w_\nu M_\mu (\sk ) 
\ee
plays the role of an effective spectrum, renormalized, as compared to the 
single - particle spectrum (42), by the presence of the mass operator (43).

The Hamiltonian (49) can be diagonalized by the Bogolubov canonical 
transformation
\be
c_s(\sk ) =u(\sk )a_s(\sk ) + v(\sk )\stackrel{\dagger}{a}_{-s}(-\sk ) ,
\ee
in which both $ \; c_s \; $ and $ \; a_s \; $ satisfy the Fermi commutation 
relations. Diagonalization is achieved with
$$ \left | u(\sk ) \right |^2 = \frac{1}{2} \left | 1 + 
\frac{\omega_1(\sk )}{E_1(\sk )} \right | , \qquad \left | v(\sk ) 
\right |^2 =
\frac{1}{2} \left | 1 - \frac{\omega_1(\sk )}{E_1(\sk )} \right | $$
leading to the Hamiltonian
\be
H_1 = w_1\sum_{k,s} E_1(\sk )\stackrel{\dagger}{a}_s(\sk )a_s(\sk ) + 
w_1C_1 ,
\ee
where the quasiparticle spectrum
\be
E_1^2(\sk ) \equiv \Delta_1^2(\sk ) + \omega_1^2(\sk )
\ee
contains the gap
\be
\Delta_1(\sk ) \equiv -\frac{w_1}{V}\sum_{p}J_1(\sk ,\spp )\sigma_1(\spp ) .
\ee
For the scalar part of (52) we have
\be
c_\nu \equiv \sum_{k} \left [ \omega_\nu (\sk ) - E_\nu (\sk ) + 
\Delta_\nu (\sk )\sigma_\nu (\sk ) -\frac{w_\nu}{2}M_\nu (\sk )n_\nu (\sk )
\right ] .
\ee

With the diagonal hamiltonian (52) it is straightforward to calculate the 
momentum distribution (34),
\be
n_1(\sk ) = 1 -\frac{\omega_1(\sk )}{E_1(\sk )} \tanh 
\frac{w_1E_1(\sk )}{2T} ,
\ee
and the anomalous averages (35),
\be
\sigma_1(\sk ) =\frac{\Delta_1(\sk )}{2E_1(\sk )}\tanh
\frac{w_1E_1(\sk )}{2T} .
\ee
So, for the gap (54) we obtain the equation
\be
\Delta_1(\sk ) = -\frac{w_1}{V}\sum_{p}J_1(\sk ,\spp ) 
\frac{\Delta_1(\spp )}{2E_1(\spp )}\tanh \frac{w_1E_1(\spp )}{2T} .
\ee

Let us emphasize that in the case of a superconductor with phase separation, 
with which we are dealing, all formulas obtained above involve the phase 
probabilities $ \; w_\nu \; $ in an intricate way. And it is very important 
where and how the latter enter into the expressions. Even though the technical
approximations we have adopted are standard, it is essential to accurately 
trace the role of the phase probabilities since it is these which distinguish 
the heterophase from pure superconductor. Also, we have to carefully take into
account all interactions since, as has been shown, the presence of repulsive 
interactions is decisive for the phase separation itself.

For the superconducting phase we need a nontrivial solution of (58). For the 
normal phase according to condition (17) instead of (56) we have
\be
n_2(\sk ) = 1 - \tanh\frac{w_2\omega_2(\sk )}{2T} = 
\frac{2}{\exp [w_2\omega_2(\sk )/T]+1} ,
\ee
and instead of (57) and (58) we have
\be
\sigma_2(\sk ) = 0 , \qquad \Delta_2(\sk ) = 0 ,
\ee
with the effective spectrum, $ \; \omega_2(\sk ) \; $, given by (50).

\section{Interaction Potentials}

It follows from the previous Section that all characteristics of the system 
can be calculated provided we know the mass operator (43) and the effective 
interaction (44). These quantities are not independent of each other since 
they are both defined, by means of (38) and (44), through the vertex (25). 
If the real - space vertex entering into the hamiltonian (19), describes an 
interaction between particles, which is, as usual, invariant with respect to 
permutation of particles and space reflections, then the momentum - space 
vertex (25) has the properties [2]:
$$ \Gamma_\nu (\sk_1,\sk_2,\sk_3,\sk_4) = 
\Gamma_\nu (\sk_2,\sk_1,\sk_4,\sk_3) =  $$
$$ = \Gamma_\nu (-\sk_1,-\sk_2,-\sk_3,-\sk_4) = 
\Gamma_\nu (\sk_1,-\sk_3,-\sk_2,\sk_4) = $$
$$ = \Gamma_\nu (\sk_3,\sk_4,\sk_1,\sk_2) . $$
With these properties, the symmetrized vertex (30) coincides with (25),
$$ \gn = \Gamma_\nu (\sk_1,\sk_2,\sk_3,\sk_4) . $$
It then follows from (38),(43) and (44) that
\be
M_\nu (\sk ) =\frac{1}{V} \sum_{p} n_\nu (\spp ) \left [ J_\nu (\sk ,\sk ) -
\frac{1}{2}J_\nu (\sk ,\spp ) \right ] .
\ee
Thus, the mass operator (43) is completely defined by the effective 
interaction (44) by the relation (61).

As we have emphasized, we have to keep in the Hamiltonian (19) all 
interactions, direct and indirect. Therefore, the vertex (25) can be 
written as a sum $ \; \Gamma_\nu = \Gamma_\nu^{dir} + \Gamma_\nu^{ind} \; $ 
of direct and indirect interactions. Hence, the effective interaction (44) is 
a sum 
\be
J_\nu (\sk ,\spp ) = J_\nu^{dir}(\sk ,\spp ) + J_\nu^{ind}(\sk ,\spp ) 
\ee
of direct and indirect interactions.

In principle, we could continue our analysis employing only the occurrence 
of the separation (62) of the effective interaction into two terms, without 
specifying their nature. For example, as interacting particles we could think 
of electrons or of atoms of $ \; ^3He \; $ or of nucleons, taking for the 
real - space direct interactions the Coulomb, Lennard - Jones or Yukawa 
potentials, respectively. The indirect interaction might be assumed to be 
mediated by a boson - exchange mechanism involving phonons or excitons or 
something else.

However, we prefer to be more concrete, so that in what follows, we shall 
think of particles as electrons whose direct interaction is described by a 
screened Coulomb potential and indirect interaction induced by phonon 
exchange. Generally, the screened Coulomb potential depends on the properties 
of a phase inside which electrons interact, since the screening is described 
by an inverse dielectric function reflecting the features of matter [54,55]. 
The simplest form of screening is obtained by using the Thomas - Fermi 
approximation which is equivalent to the static approximation for the 
Lindhard dielectric function [54,55]. Then, in real space the screened 
Coulomb interaction of particles immersed in the $ \; \nu \; $- phase 
has the form
\be
\Phi_\nu (r) =\frac{e^2}{r}e^{-\kappa_\nu r} ,
\ee
in which $ \; e \; $ is the charge, $ \; \kappa_\nu^{-1} \; $ is a screening 
radius with
$$ \kappa_\nu^2 = \frac{4}{a_B}\left (\frac{3}{\pi}n_\nu \right )^{1/3} 
\qquad \left ( a_B \equiv \frac{1}{me^2} \right ) , $$
where $ \; m \; $ is the electron mass and $ \; n_\nu \; $, the electron 
density in the $ \; \nu \; $- phase. In momentum space, (63) yields
\be
v_\nu (k) = \frac{4\pi e^2}{k^2 +\kappa_\nu^2} .
\ee
Recall that the particular form (64) is not of great importance for our 
argument: for example, we could take for $ \; v_\nu (k) \; $ the more 
general expression $ (4\pi /k^2)\ep_\nu^{-1}(k,\omega_\nu^{exc}(k)) \; $, 
in which $ \; \ep_\nu (k,\omega ) \; $ is a dielectric function and 
$ \; \omega_\nu^{exc}(k) \; $, a spectrum of elementary excitations in the 
$ \; \nu \; $- phase. Also, for the dielectric function we could invoke a 
more refined approximation [55-60]. Here we have chosen the simple form (64) 
in order to illustrate how the screening can depend on the phase properties. 
In any case, whether $ \; v_\nu (k) \; $ is given by (64) or by a more 
complicated expression, the part of (62) related to the direct interaction 
is 
\be
J_\nu^{dir}(\sk ) = \int \Phi_\nu (r)e^{-i(\sk - \spp )\srr}d\srr =
v_\nu \left (  | \sk - \spp | \right ) .
\ee 
For the effective indirect interaction we take one mediated by phonon exchange
in the form [55]
\be
J_\nu^{ind}(\sk ,\spp ) = -\frac{|\al_\nu |^2}{\omega_{\nu 0}^2 - 
[\omega_\nu (\sk ) - \omega_\nu (\spp )]^2} ,
\ee
where $ \; \omega_{\nu 0} \; $ is a characteristic phonon frequency, 
$ \; \alpha_\nu \; $ is the electron - phonon coupling taking account of 
screening,
$$ \al_\nu = -\frac{4\pi i Ze^2}{k_{\nu F}(1 + \kappa_\nu^2/k_{\nu F}^2)}
\left ( \frac{\rho_\nu}{M} \right )^{1/2} , $$
$ \; Z \; $ and $ \; M \; $ are an ion charge and mass, respectively, 
$ \; \rho_\nu \; $ is the density of ions in the $ \; \nu \; $- phase, and 
$ \; k_{\nu F} \approx (3\pi^2n_\nu )^{1/3} $ is the Fermi momentum of an 
electron.

The presentation (62) of the effective interaction as a sum of direct and 
indirect interactions leads to a similar decomposition
\be
M_\nu (\sk ) = M_\mu^{dir}(\sk ) + M_\nu^{ind}(\sk ) 
\ee
of the mass operator (61). The first term in (67) is related to (65) giving
\be
M_\nu^{dir}(\sk ) = n_\nu v_\nu (0) - \frac{1}{2V} \sum_{p}n_\nu (\spp )
v_\nu ( | \sk - \spp | ) ,
\ee
while the second term in (67) is expressed through (66) yielding
\be
M_\nu^{ind}(\sk ) = -n_\nu \frac{|\al_\nu |^2}{\omega_{\nu 0}^2} -
\frac{1}{2V}\sum_{p} n_\nu (\spp )J_\nu^{ind}(\sk ,\spp ) ,
\ee
where
\be
n_\nu \equiv \frac{1}{V}\sum_{p}n_\nu (\spp )
\ee
is the electron density in the $ \; \nu \; $- phase.

The chemical potential, as a function of temperature and density, is defined 
by the formula
\be
N = - \langle\frac{\pr \stackrel{-}{H}}{\pr \mu} \rangle =
\sum_{\nu} w_\nu \int \langle \stackrel{\dagger}{\psi}_\nu (\srr )
\psi_\nu (\srr )\rangle d\srr 
\ee  
for the total number of electrons in the system. Eq.(71), with notation (34) 
and (36), gives
\be
N = \sum_{p} \left [ w_1n_1(\spp ) + w_2n_2(\spp ) \right ] .
\ee
Using (70), we can reduce (72) to the equation
\be
n \equiv \frac{N}{V} = w_1n_1 + w_2n_2
\ee
connecting the average electron density in the system with the electron 
densities in each of the phases which the system is composed of and with 
the corresponding phase probabilities. Note that when the electron densities 
in both phases coincide with each other, that is when $ \; n_\nu = n \; $, 
then (73) reduces to (7).

Considering the gap equation (58) we pass to the thermodynamic limit with the 
standard replacement $\;\frac{1}{V}\sum_{p}\ra\int\frac{d\spp}{(2\pi )^3}\;$. 
The main contribution to the resulting integral in the right - hand side of 
(58) comes from momenta close to the Fermi surface which is defined by the 
condition
\be
\omega_1(\sk_F) = 0 .
\ee
Keeping this in mind, we can rewrite the gap equation (58) as
$$ \Delta_1(\sk ) = \frac{w_1}{2} \left [ 
\frac{|\al_1|^2}{\omega_{10}^2 - \omega_1^2(\sk )} - \stackrel{-}{v}_1(\sk )
\right ] \times $$
\be
\times \int \frac{\Delta_1(\spp )}{\sqrt{\Delta_1^2(\spp ) + 
\omega_1^2(\spp )}}
\tanh \frac{w_1\sqrt{\Delta_1^2(\spp )+ \omega_1^2(\spp )}}{2T}
\cdot \frac{d\spp }{(2\pi )^3} ,
\ee
where
$$ \stackrel{-}{v}_1(k) \equiv \lim_{p \ra k_F} \frac{1}{4\pi} 
\int v_1(|\sk - \spp |)d\Omega (\spp ) = $$
\be
= \frac{\pi e^2}{kk_F}\ln \left | 
\frac{(k + k_F)^2 + \kappa_1^2}{(k - k_F)^2 + \kappa_1^2} \right |
\ee
is the screened Coulomb interaction averaged over spherical angles.

A necessary condition that (75) has a nonzero solution for the gap 
$ \; \Delta_1(\spp ) \; $ is
\be
\frac{|\al_1|^2}{\omega_{10}^2 - \omega_1^2(\sk )} - 
\stackrel{-}{v}_1(k) > 0 ,
\ee
which occurs in the vicinity of the Fermi surface so that 
$ \; \omega_1^2(\sk )< \omega_{10}^2 \; $. Actually, inequality (77) has 
almost the same form as the BCS criterion for superconductivity [1]. However 
in our case the quantities entering into (77) depend on the superconducting 
phase probability. This dependence is essential and as we show below can 
dramatically change the characteristics of superconductors with phase 
separation making the existence of superconductivity possible even for values 
of the parameters which in a pure sample would imply the absence of 
superconductivity.

Equation (75) can be simplified by considering the value of the gap at the 
Fermi surface, i.e.
$$ \Delta \equiv \lim_{k \ra k_F} \Delta_1(\sk ) , $$
 and by introducing the level density per spin,
\be
N_1(\omega ) \equiv \frac{1}{(2\pi )^3} \int 
\frac{dS(\omega )}{|\nabla_k \omega_1(\sk ) |_\omega } ,
\ee
where the integration is over the surface given by the equation 
$ \; \omega_1 (\sk ) = 0 \; $. Then (75) yields
\be
\int_{0}^{\omega_{10}}\frac{\lambda_{eff}}{\sqrt{\Delta^2+\omega^2}}
\tanh \frac{w\sqrt{\Delta^2+\omega^2}}{2T}d\omega = 1 ,
\ee
where the effective coupling parameter is
\be
\lambda_{eff} \equiv wN_1(0) \left [ \frac{|\al_1|^2}{\omega_{10}^2} - 
\stackrel{-}{v}_1(k_F) \right ] ,
\ee
$ \; N_1(0) \; $ being the level density (78) at the Fermi surface defined 
by (74).

The criterion (77) implies that for superconductivity the effective coupling 
parameter (80) is positive. For metals one has $ \; \kappa_1 \sim k_F \; $ 
and $ \; \stackrel{-}{v}_1(k_F) \sim \pi e^2/k_F^2 \; $. Therefore in 
this case we have the estimate
$$ \frac{\stackrel{-}{v}_1(k_F)}{|\al_1|^2/\omega_{10}^2} \sim 
\frac{\omega_{10}^2}{\Omega_p^2} \qquad (\Omega_p^2 \equiv 
\frac{4\pi}{M}\rho_1Z^2e^2 ) , $$
in which $ \; \Omega_p \; $ is the ion plasma frequency. The condition 
$ \; \lambda_{eff} > 0 \; $ means that $ \; \omega_{10} < \Omega_p \; $. 
The latter inequality makes it possible to understand why structural 
instabilities and the related softening of the lattice, typical of 
high - temperature superconductors, favour the appearance of 
superconductivity. Lattice softening is associated with a decrease in the 
characteristic phonon frequency $ \; \omega_{10} \; $, this makes it easy 
to satisfy the condition $ \; \omega_{10} < \Omega_p \; $ and favours the 
onset of superconductivity. However, the latter occurs not in the whole 
volume of a sample, but only in those parts of it that are occupied by 
the superconducting phase. This is why lattice softening, enhancement 
of superconductivity and phase separation are phenomena that are intimately 
related to one another. This  conclusion is supported as well by the 
consideration of s Hubbard - type model for which it has been found [42] 
that superconducting correlations are enhanced in the phase - separation 
regions.

\section{Critical Temperature} 

The superconducting phase transition occurs at the critical temperature 
$ \; T_c \; $ where $ \; \Delta = 0. \; $. The equation for $ \; T_c \; $ 
follows from (79):
\be
\lambda_{eff}\int_{0}^{\omega_{10}}\frac{1}{\omega}\tanh 
\frac{w\omega}{2T_c}d\omega = 1 .
\ee
Recall that here $ \; w(T_c) \; $ is the geometric probability of the 
superconducting phase, that is, the relative part of volume occupied by this 
phase. The upper limit in the integral (81) is the characteristic phonon 
frequency exhibiting softening because of structural fluctuations and the 
related phase separation [18,61]. The simultaneous occurrence of lattice 
softening and phase separation can be clearly observed by using M\"ossbauer 
spectroscopy [62,63]. In high - temperature superconductors these effects 
lead to anomalous saggings of the M\"ossbauer factor at critical temperature 
[12,13,64]. The softening of phonon frequencies at $ \; T_c \; $ can also be 
observed by other experimental methods such as infrared, Raman and neutron 
scattering [11]. The softening of the characteristic frequency 
$ \; \omega_{10} \; $, as was shown in [61,63], can be expressed by the 
relation
\be
\omega_{10} =w^\varphi\omega_0 ,
\ee in which $ \; \omega_0 \; $ is a characteristic phonon frequency of a 
pure superconductor, without phase separation, and the parameter $\;\varphi\;$
measures the intensity of softening. Weak softening corresponds to 
$ \; \varphi = 1/2 \; $, moderate, to $ \; \varphi = 1 \; $, and the strong 
softening, to $ \; \varphi =3/2 \; $.

For superconductors with phase separation the dependence of the critical 
temperature $ \; T_c \; $ on the superconducting - phase probability 
$ \; w \; $ is given by an intricate relation through (80) - (82). If one 
interprets the onset of superconductivity as the condensation of Cooper pairs,
then one can call $ \; w \; $ the concentration of superconducting condensate.
No matter what it is called, the main point is that this quantity can 
be varied in high - temperature superconductors by varying their chemical 
structure, for example by doping. The experimentally measured dependence of 
$ \; T_c \; $ on $ \; w \; $ exhibits, for some high - temperature 
superconductors, a quite unusual nonmonotonic behaviour, as is reviewed in 
refs.[65,66]. For this reason it is especially interesting to study the 
dependence of $ \; T_c \; $ on $ \; w \; $.

Formula (81) allows us to obtain two asymptotic expressions. One limit 
corresponds to the case
\be
T_c \ll \frac{\omega_0}{\pi}w^{1+\varphi} , \qquad \lambda_{eff} \ll 1 ,
\ee
when
\be
T_c \simeq 1.134w^{1+\varphi}\omega_0\exp \left ( -\frac{1}{\lambda_{eff}}
\right ) .
\ee   
Note, that if we consider $ \; \omega_{eff} \equiv w^{1+\varphi}\omega_0 \; $ 
and $ \; \lambda_{eff} \; $ as fitting parameters, then expression (84), as 
is known [67,68], describes the majority of low - temperature superconductors 
quite well.

Another limiting case, opposite to (83), is when
\be
T_c \gg \frac{\omega_0}{\pi}w^{1+\varphi}, \qquad \lambda_{eff} \gg 1 .
\ee
Then (81) gives
\be
T_c \simeq \frac{1}{2}w^{1+\varphi}\lambda_{eff}\omega_0 .
\ee
There is a temptation to treat (85) as the strong coupling limit. In doing 
this, we should be very cautions and not forget that $ \; \lambda_{eff} \; $, 
given by (80), is an effective coupling parameter. So, it may happen that 
$ \; \lambda_{eff} \; $ is large, even if the bare coupling constant, 
$ \; \lambda \; $ is not. If $ \; \lambda \gg 1 \; $, then by a standard 
argument Eliashberg equations imply that $ \; T_c \sim \sqrt{\lambda} \; $, 
although the dependence $ \; T_c \sim \lambda \; $, as is stated in [69], is 
also consistent with the strong coupling limit of these equations.

The involvement of the superconducting phase concentration in the definition 
of the effective coupling parameter (80) makes the applicability of such 
simple asymptotic expressions, as (84) and (86), to high - temperature 
superconductors with phase separation quite limited.

We therefore attempt an accurate analysis of the dependence of $ \; T_c \; $ 
on $ \; w \; $. First, we have to remember that the level density (78) also 
depends on $ \; w \; $ through the effective spectrum (50) renormalized 
by the mass operator (67). Consider the isotropic case leaving aside the 
very interesting, but separate problem of van Hove singularities [7]. Then, 
differentiating the effective spectrum (50), we find
\be
\lim_{k \ra k_F} \left | \nabla_k\omega_1(\sk ) \right | = 
\ep_F' + wM_F' ,
\ee
where
$$ \ep_F' \equiv \lim_{k \ra k_f}\frac{\pr}{\pr k}\ep_1(k) , \qquad
M_F' \equiv \lim_{k \ra k_F}\frac{\pr}{\pr k}M_1(k) . $$
The level density (78) at the Fermi surface becomes
\be
N_1(0) =\frac{N_1^*(0)}{|1 +\gamma w|} ,
\ee
where we use the notation
\be
N_1^*(0) \equiv \frac{k_F^2}{2\pi^2|\ep_F'|} , \qquad \gamma \equiv
\frac{M_F'}{\ep_F'} .
\ee
To estimate the value of $ \; M_F' \; $, take into account that 
$ \; n_1 \approx k_F^3/3\pi^2 \; $ and $ \; \kappa_1 \approx k_F \; $, then
$ \;  M_F' \approx 4e^2/3\pi \; $. For a parabolic zone $ \; \ep_F' \approx
k_F/m \; $, so that for the parameter $ \; \gamma \; $ in (89) we get 
$ \; \gamma \sim a_e/\pi^2a_B \; $, where $ \; a_e \; $ is the average 
distance between electrons, and $ \; a_B \; $ is the Bohr radius. In good 
conductors $ \; a_e \sim a_B \; $, whence $ \; \gamma \sim 0.1 \; $. This 
means that in good metals the renormalization of the spectrum, due to the 
mass operator, is quite weak.

In bad conductors with low electron density one has $ \; a_e \gg a_B \; $. 
Therefore $ \; \gamma \; $ may equal $ \; 1 \; $ or more. This makes the role 
of the mass operator in renormalizing the electron spectrum and consequently 
in influencing the level density (88) very important. Such a situation is 
directly related to high - temperature superconductors which are, as is known,
bad conductors having a low density of carriers.

To distinguish the effects due to phase separation, we introduce the 
quantities
\be
\lambda^* \equiv N_1^*(0)\frac{|\al_1|^2}{\omega_0^2} , \qquad
\mu^* \equiv N_1^*(0)\stackrel{-}{v}_1(k_F) 
\ee
which do not depend on the phase probability $ \; w \; $. The first quantity 
in (90) is the electron - phonon coupling constant for a {\it pure} system 
without phase separation. The second quantity in (90) is the average intensity
of screened Coulomb interaction.

With notation (90), for the effective coupling parameter (80) we obtain
\be
\lambda_{eff} =
\frac{\lambda^* -\mu^*w^{2\varphi}}{w^{2\varphi -1}|1 +\gamma w |} .
\ee
The criterion of superconductivity means that $ \; \lambda_{eff} > 0 \; $, 
which leads because of (91) to
$$ \lambda^* - \mu^*w^{2\varphi} > 0 . $$
This inequality, in the case of pure superconductor, when $\; w\equiv 1\;$, 
reduces to the standard condition $ \; \lambda^* > \mu^* \; $ usually valid 
for low - temperature superconductors. When the phase separation occurs in a 
superconductor, then the condition $ \; \lambda_{eff} > 0 \; $ is easier to 
satisfy. Really, when $ \; 0 < w < 1 \; $, then the inequality 
$ \; \lambda^* >\mu^*w^{2\varphi} \; $ can be true even if 
$ \; \lambda^* < \mu^* \; $. In this way, phase separation favours the 
appearance of superconductivity. A sample with phase separation can become 
superconductive even if a similar sample, without phase separation, cannot 
have superconducting properties.

The superconducting critical temperature is given by (81) which becomes
\be
\frac{\lambda^* -\mu^*w^{2\varphi}}{w^{2\varphi -1}|1+\gamma w|}
\int_{0}^{w^\varphi\omega_0} \frac{1}{\omega} \tanh \frac{w\omega}{2T_c} 
d\omega = 1 .
\ee
As we see, it is not easy to analyse precisely the influence of phase 
separation on the critical temperature, that is, to solve explicitly for the 
dependence of 
$ \; T_c \; $ on the phase probability $ \; w \; $. To slightly simplify the 
situation, we may notice that since $ \; \ep_F' \approx k_F/m , \; N_1^*(0) 
\approx mk_F/2\pi^2 \; $ and $ \; \stackrel{-}{v}_1(k_F) \approx 
\pi e^2/k_F^2 \; $, then $ \; \mu^* \approx \gamma \; $. Therefore, for 
simplicity, we put $ \gamma = \mu^* \; $.   

Of course, such a slight simplification does not help much, and to proceed 
further in defining the dependence of $ \; T_c \; $ on $ \; w \; $ we have 
to resort to numerical analysis of (92). To this end, it is convenient to 
introduce the dimensionless critical temperature
$$ t_c \equiv \frac{T_c}{\omega_0} . $$
Now we can reorganize (92) in the form 
\be
\frac{\lambda^* -\mu^*w^{2\varphi}}{w^{2\varphi -1}(1+\mu^*w)}
\int_{0}^{1}\tanh \left ( \frac{w^{1+\varphi}}{2t_c}x\right )
\frac{dx}{x} = 1 .
\ee
From (93) we can immediately conclude that the behaviour of 
$ \; t_c = t_c(w) \; $ can be, in general, nonmonotonic, since there are 
two points at which $ \; t_c(w) \; $ tends to zero. The first point 
corresponds to $ \; w \ra 0 \; $. Then from (93) we obtain
\be
t_c \simeq \frac{1}{2}w^{2-\varphi}
\frac{\lambda^* -\mu^*w^{2\varphi}}{1+\mu^*w} .
\ee
The second case, when $ \; t_c \ra 0 \; $, is when 
$ \; w \ra (\lambda^*/\mu^*)^{1/2\varphi} \; $, then
\be
t_c \simeq 1.134w^{1+\varphi}\exp 
\left \{ - \frac{w^{2\varphi -1}(1+\mu^*w)}{\lambda^* -\mu^*w^{2\varphi}} 
\right \} .
\ee
Recall that the dependence of $ \; t_c \; $ on $ \; w \; $ is interesting 
to analyse because the superconducting phase concentration $ \; w \; $ can 
be measured and can be varied in experiments by changing the chemical 
structure of materials, for example, by doping [6,11,65,66].

We made a detailed analysis of the function $ \; t_c(w) \; $ by solving the 
equation (93) numerically. Graphs of the resulting functions are presented 
in figs.1-12. We did not try to fit any particular experimental situation. 
Rather we wanted to understand the whole picture as to how the behaviour of 
$ \; t_c(w) \; $ changes qualitatively with the change of parameters 
$ \; \varphi ,\mu^* \; $ and $ \; \lambda^* \; $.

We were pleasantly surprised by the wide variety of curves which were 
obtained. It is apparent from the accompanying graphs that by the choice 
of the corresponding parameters it will be possible to obtain a reasonably 
good fit to any experimental curve.

Figs.1-4 correspond to the case of weak softening; figs.5-8, to that of 
moderate softening; and figs.9-12, to the case of strong softening. With 
the increase of the Coulomb parameter the function $ \; t_c(w) \; $ really 
becomes nonmonotonic. The behaviour of $ \; t_c(w) \; $ plotted in figs.3,4,7 
and 8 has a striking similarity to the corresponding experimental curves for 
high - temperature cuprate superconductors (see [6,11,65,70] and references 
therein). The coincidence becomes practically complete if we redraw our 
figures in the relative coordinates $ \; \stackrel{-}{T} 
\equiv t_c/t_c^{max} , \; \sigma \equiv w/w^{max} \; $, as in [65,66], 
where the point $ \; (t_c^{max},w^{max}) \; $ denotes the point of a maximum 
of the considered curve.

It is clear from the preceeding analysis, as illustrated in our graphs, that,
for fixed $ \; \varphi ,\; \mu \; $ and $ \; \lambda\;$, the value of the 
reduced critical temperature, $ \; T_c \;$, depends crucially on the parameter
$ \; w\;$. The operational meaning of this parameter is rather different 
according as the the system is 1) stable, or 2) metastable.

Notice that while $\; w \;$ occurs explicitly on the left hand side of 
equation (12) it also occurs implicitly in a complicated manner in the 
right hand side as a result of the renormalization of the Hamiltonian 
(1). Thus for a stable system satisfying conditions (9) and (13), we can 
think of $\; w \;$ as determined self - consistently  by equation (12) 
once all characteristics of the system, - such as chemical composition, 
particle masses, interaction potentials, temperature, density and 
pressure - are given. Note that all of these characteristics are necessary 
for $\; w \;$ to be uniquely determined. In particular, the total role of 
interactions must be taken into account. In contrast to this, following 
common practice, in our model the gap equation (79) and consequently equation
(93) for the critical temperature, take into account only interactions 
specified on the Fermi surface. Essential to the argument of the present 
paper is the conviction that it would not be reasonable to try to treat 
superconductivity in heterophase system with a model in which $\; w\;$ is 
determined by parameters defined merely on the Fermi surface. For instance, 
parameters characterizing the ground state would be indispensable. The 
equations for phase probabilities always contain more characteristic constants
than the equations for an order parameter [18]. It is therefore permissible to
contemplate the possibility of holding some parameters, such as 
$\; \lambda^*\;$ and $\; \mu^*\;$, fixed while $\; w\;$ varies as a 
result of changing other parameters such as the chemical composition.

On the other hand, for a metastable system for which (13) does not hold even 
though thermal and mechanical stability are preserved, the fraction $\; w\;$
is not necessarily determined by equation (12), but might be arbitrary 
depending on the preparation of the sample. This possibility should not be 
overlooked since many high -$\; T_c\;$ superconductors are metastable.

\section{Conclusion}

We have developed an approach, for describing superconductors, taking into 
account three mutually interrelated factors: (i) repulsive interactions, (ii) 
lattice softening and (iii) phase separation. We have deliberately limited 
ourselves to the use of only commonly accepted approximations. This was in 
order to emphasize that our results are not artifacts of some technical 
tricks, but the direct consequences of the physical reasons offered. The 
main results can be summarized as follows:

(i) A necessary condition for phase separation in a superconductor is the 
presence of repulsive interactions.

(ii) Phase separation favours superconductivity making it possible in a 
heterophase sample even if it were impossible in a pure sample.

(iii) The superconducting critical temperature as a function of the 
relative concentration of the superconducting phase can display the 
nonmonotonic behaviour typical of high - temperature cuprate superconductors.

It should be a straightforward matter to adapt the basic approach of this 
paper to other models of superconductors by using alternative approximation 
methods.

\vspace{1cm}

{\bf Acknowledgement}

\vspace{0.5cm}

We appreciate financial support by the Natural Sciences and Engineering 
Research Council of Canada.

\newpage

\begin{center}
{\bf Figure Captions}
\end{center}

\vspace{1cm}

{\bf Fig.1}. 

The superconducting critical temperature as a function of the superconducting 
phase concentration for the parameters $ \; \varphi =0.5 , \; \mu^* = 0.1 \;$ 
and $ \; \lambda^* = 1 \; $ (lower curve), $ \; \lambda^* = 5 \; $ (middle 
curve) and $ \; \lambda^* = 10 \; $ (upper curve).

\vspace{1cm}

{\bf Fig.2}. 

The same as in fig.1, but for the parameters $\;\varphi =0.5,\;\mu^*=1 \; $, 
and $ \; \lambda* = 1 \; $ (lower curve), $ \; \lambda^*=5 \; $ (middle curve)
and $ \; \lambda^* = 10 \; $ (upper curve).

\vspace{1cm}

{\bf Fig.3}. 

The same as in fig.1 for the parameters $ \; \varphi =0.5, \; \mu^*=5 \; $, 
and $ \; \lambda* = 1 \; $ (hardly visible lower curve), $ \; \lambda^*=5 \; $
(middle curve) and $ \; \lambda^* = 10 \; $ (upper curve).

\vspace{1cm}

{\bf Fig.4}. 

The same as in fig.1 for the parameters $ \; \varphi =0.5, \; \mu^*=10 \; $, 
and $ \; \lambda* = 5 \; $ (lower curve) and $ \; \lambda^* = 10 \; $ (upper 
curve). The curve corresponding to $ \; \lambda^* = 1 \; $ in this case is 
invisible.

\newpage

{\bf Fig.5}. 

The same as in fig.1 for the parameters $ \; \varphi = 1, \; \mu^*=0.1 \; $, 
and $ \; \lambda* = 1 \; $ (lower curve), $ \; \lambda^*=5 \; $ (middle curve)
and $ \; \lambda^* = 10 \; $ (upper curve).

\vspace{1cm}

{\bf Fig.6}. 

The same as in fig.1 for the parameters $ \; \varphi =1, \; \mu^*=1 \; $, and 
$ \; \lambda* = 1 \; $ (lower curve), $ \; \lambda^*=5 \; $ (middle curve) and
$ \; \lambda^* = 10 \; $ (upper curve).

\vspace{1cm}

{\bf Fig.7}. 

The same for the parameters $ \; \varphi =1, \; \mu^*=5 \; $, and 
$ \; \lambda* = 1 \; $ (lower curve), $ \; \lambda^*=5 \; $ (middle 
curve) and $ \; \lambda^* = 10 \; $ (upper curve).

\vspace{1cm}

{\bf Fig.8}. 

The same for the parameters $ \; \varphi =1, \; \mu^*=10 \; $, and 
$ \; \lambda* = 1 \; $ (lower curve), $ \; \lambda^*=5 \; $ (middle 
curve) and $ \; \lambda^* = 10 \; $ (upper curve).

\vspace{1cm}

{\bf Fig.9}. 

The same for the parameters $ \; \varphi =1.5, \; \mu^*=0.1 \; $, and 
$ \; \lambda* = 1 \; $ (lower curve), $ \; \lambda^*=5 \; $ (middle curve) 
and $ \; \lambda^* = 10 \; $ (upper curve).

\vspace{1cm}

{\bf Fig.10}. 

The same for the parameters $ \; \varphi =1.5, \; \mu^*=1 \; $, and 
$ \; \lambda* = 1 \; $ (lower curve), $ \; \lambda^*=5 \; $ (middle curve) 
and $ \; \lambda^* = 10 \; $ (upper curve).

\vspace{1cm}

{\bf Fig.11}. 

The same for the parameters $ \; \varphi =1.5, \; \mu^*=5 \; $, and 
$ \; \lambda* = 1 \; $ (lower curve), $ \; \lambda^*=5 \; $ (middle 
curve) and $ \; \lambda^* = 10 \; $ (upper curve).

\vspace{1cm}

{\bf Fig.12}. 

The same for the parameters $ \; \varphi =1.5, \; \mu^*=10 \; $, and 
$ \; \lambda* = 1 \; $ (lower curve), $ \; \lambda^*=5 \; $ (middle 
curve) and $ \; \lambda^* = 10 \; $ (upper curve).

\end{document}